\documentclass[10pt,preprint]{aastex}
\usepackage{graphicx} 
\usepackage{color}

\shortauthors{Alessandrini et al.}

\def\ltsima{$\; \buildrel < \over \sim \;$}
\def\gtsima{$\; \buildrel > \over \sim \;$}
\def\lsim{\lower.5ex\hbox{\ltsima}}
\def\gsim{\lower.5ex\hbox{\gtsima}}

\begin{document}

\title{Investigating mass segregation process in globular clusters
  with Blue Straggler Stars: the impact of dark remnants}

\author{Emiliano Alessandrini$^1$,
Barbara Lanzoni$^1$,
Francesco R. Ferraro$^1$,
Paolo Miocchi$^1$,
Enrico Vesperini$^2$}

\affil{\textsuperscript{1} Dept. of Physics and Astronomy, University of
Bologna, viale Berti Pichat, 6/2.
\textsuperscript{2} Dept. of Astronomy, Indiana University,
Bloomington, IN, 47401, USA.}

\date{19 sept 2016}

\begin{abstract}
We present the results of a set of N-body simulations aimed at
exploring how the process of mass segregation (as traced by the
spatial distribution of blue straggler stars, BSSs) is affected by the
presence of a population of heavy dark remnants (as neutron stars and
black holes). To this end, clusters characterized by different initial
concentrations and different fractions of dark remnants have been
modeled. We find that an increasing fraction of stellar-mass
  black holes significantly delays the mass segregation of BSSs and
the visible stellar component.  In order to trace the evolution of
BSS segregation, we introduce a new parameter ($A^+$) that can be
easily measured when the cumulative radial distribution of these stars
and a reference population are available. Our simulations show
  that $A^+$ might also be used as an approximate indicator of the
  time remaining to the core collapse of the visible component.
\end{abstract}


\section{Introduction}
\label{intro}
Mass segregation is one of the manifestations of two-body relaxation
in collisional stellar systems.  Kinetic energy exchanges between
stars with different masses cause a slowing down of the heaviest one,
which spirals toward the centre of the system in favour of a velocity
enhancement of the lighter one, which, instead, migrates toward the
outskirt.  Since the time scale of this process is the dynamical
friction (DF) timescale $t_{\rm DF}$ \citep{chandrasekhar43}, which
depends on the local velocity dispersion and (especially) on the local
density of the cluster \citep[e.g.,][]{binney+tremaine,
  alessandrini+14}, it is natural to expect different stellar systems
to show different levels of mass segregation.

In this respect, Blue Straggler Stars (BSSs) have been used as
powerful observational test particles to probe mass segregation in
globular clusters (GCs; e.g., \citealp{mapelli+04, mapelli+06,
  ferraro+93, ferraro+97, ferraro+04, ferraro+06b, lanzoni+07a,
  lanzoni+07b, dalessandro+08, beccari+11}).  This is because BSSs in
GCs have a mass of $\sim 1.2-1.4 M_\odot$ (\citealp{shara97,
  ferraro+06a, lanzoni+07a, fiorentino+14}) which is larger than the
main-sequence turn-off (MS-TO) mass and the average stellar mass in
the system ($0.8 M_\odot$ and $\sim 0.5 M_\odot$, respectively).
Hence, BSSs experience significant DF against "normal'' cluster
stars. Moreover, these objects are relatively bright, with visual
magnitudes up to 2.5 dex above the MS-TO point
\citep[e.g.,][]{sandage53, ferraro+93, lanzoni+07a, dalessandro+12},
and they can be efficiently distinguished from the normal low-mass
stars in the cluster.  Indeed, previous works largely demonstrated
that BSSs tend to be more centrally concentrated than normal cluster
populations (e.g., \citealp{ferraro+99}), and different levels of
segregation have been found in different GCs (see, e.g., Figure 2 in
\citealp{ferraro+03}).  Recently \citet{ferraro+12} proposed to use
the shape of the normalized BSS radial distribution\footnote{With
  "normalized BSS distribution'' here we indicate the double
  normalized ratio defined in \citet{ferraro+93}} as an indicator of
the level of dynamical evolution experienced by a stellar system.

Following this approach GCs can be grouped in three main
families ({\it Family I, II and III}) of increasing dynamical ages on
the basis of the different shapes of their BSS radial distribution.  In the
scenario proposed by \citet{ferraro+12}, DF is the main phenomenon
that drives the BSS sedimentation toward the center of the
system. Hence, {\it Family I GCs}, where BSSs follow the same radial
distribution of the reference population, are dynamically young
systems, where DF has not started yet to segregate these heavy
objects, not even those orbiting the most central regions. In {\it
  Family II GCs}, where the BSS radial distribution is bimodal with a
central peak, a minimum in the intermediate region, followed by a
rising branch in the external part, the action of DF progressively
affected BSSs at increasingly larger distances from the centre: these
systems are dynamically intermediate-age GCs. In particular, four
different {\it sub-Families II}, with progressively increasing
dynamical ages according to the increasing radial position of the
minimum of their BSS distribution, have been defined by
\citeauthor{ferraro+12} (2012; see their Figure 2). Finally, in {\it
  Family III GCs}, where the BSS radial distribution shows only the central peak
and then monotonically decreases at increasing distance from the
cluster center, DF nominally segregated the entire population of BSSs:
these are dynamically-old GCs. 

Theoretical confirmations of this result have been searched by means
of semi-analytic and numerical approaches. \citet{alessandrini+14}
excluded the possibilty that the shape of the BSS radial distribution
in Family II clusters is due to a combination of different DF times
for different mass groups in a multi-mass system, thus confirming that
the scenario proposed in \citet{ferraro+12} is the most likely
explanation of the observed shapes of the BSS distribution.  The recent
numerical work by \citet{miocchi+15} provided some additional support to the
observations, confirming the formation of a sharp central peak, which
remains a stable feature over time, regardless of the initial
concentration of the system.  In spite of a noisy behaviour, a bimodal
distribution is seen in several cases, and in the most advanced
stages, the distribution becomes monotonic. However, in that work, the
minimum of the BSS distribution is not always easily identifiable, and
its outward migration occurs over a very short timescale, thus
suggesting an insufficient level of realism of the adopted
models. Admittedly, in \citet{miocchi+15} the total number of particles
was limited to only $N=10^4$, and the mass spectrum was roughly
modelled with only three mass bins. Moreover no populations of white
dwarfs (WDs) and dark remnants (DRs), composed of neutron stars (NSs)
and stellar mass black holes (BHs) were included. The dark
component (BHs and NSs) is thought to substantially influence the
cluster dynamics and structural
properties \citep{sigurdsson+hernquist93, mackey+07, mackey+08,
  morscher+15}. In fact, since their masses are significantly larger
than the average, DRs are thought to be the main drivers of the
cluster core-collapse (CC) phase, forming close binaries that strongly
interact dynamically with the other stars approaching the core during
the evolution.

Hence, for the proper interpretation of the observational results,
more realistic models of GCs are needed. In the present paper we
discuss a set of simulations that, with respect to those in
\citet{miocchi+15}, include $(i)$ a significantly larger number ($N
\simeq 10^5$) of particles, $(ii)$ a much finer mass-spectrum obtained
from the evolution of a \citet{kroupa01} Initial Mass
Function (IMF), and $(iii)$ a population of DRs.  The number of DRs in
present-day GCs is poorly constrained and, during the last 30 years,
many authors have addressed the problem of how many NSs
\citep[e.g.][]{hut+91, drukier96, davies+hansen98, pfahl+02,
  podsiadlowski+05} and BHs \citep[e.g.,][]{sigurdsson+hernquist93,
  oleary+06, moody+sigurdsson09, repetto+12, morscher+13,
  sippel+hurley13} are retained during the evolution of the cluster,
in what is called the {\itshape{retention problem}}. For this reason,
we performed simulations with different fractions of NSs and BHs. We
emphasize that the models presented here are still idealized and
aimed at exploring some fundamental aspects of the mass segregation
process; additional ingredients (such as a population of primordial
binaries or the effect of an external tidal field) are needed to
approach more realistic models of GCs, and they will be the goal of
forthcoming papers.
By taking into account the limitations that are present also in the
current approach, instead of following the evolution of the shape of
the BSS distribution, here we focus on the definition of a new
parameter able to quantify the level of sedimentation of BSSs toward
the cluster center. The paper is organized as follows: in
Section \ref{Nbody} we present our set of N-body models, describing
the initial conditions of our simulations; in Sections
\ref{lagr}--\ref{cumu} we discuss the time evolution of the Lagrangian
radii and the cumulative radial distributions of BSSs
and reference stars; in Sections \ref{a+} we define a new
mass segregation indicator and discuss its time dependence. Section
\ref{conclusions} summarizes the results obtained and
future perspectives.

\section{$N$-body models}
\label{Nbody}
The simulations exploit the Graphic Processing Unit (GPU) version of
the direct $N$-body code {\tt{NBODY6}} \citep{nitadori+aarseth12} on
the {\tt{BIGRED2}} supercomputer at Indiana University,
Bloomington. We performed ten different runs in order to explore
both the effect of various percentages of DRs, and the effect of
different cluster concentrations, on GC dynamical evolution.

The initial conditions have been built as follows. Starting from
$99700$ stars belonging to a \citet{kroupa01} IMF, in the mass
interval $m=[0.1,100] M_\odot$ and assuming a metallicity $Z = 0.001$,
we evolved the system for 12 Gyr, by means of the stellar evolution
recipes implemented in the SSE version of the software {\tt{McLuster}}
\citep{hurley+00, hurley+02,kupper+11}.  This procedure generated a
population of WDs and DRs descending from the evolution of stars with
initial masses $m>0.8 M_\odot$. In particular, NSs have masses
  peaked at $1.4 M_\odot$, with a tail up to $2.5 M_\odot$, while BH
  masses range between $2.5 M_\odot$ and $\sim 25 M_\odot$.  To all
the runs we also added $300$ BSSs, modeled as single particles with a
mass of $1.2 M_\odot$.  The number of BSSs in our runs, although being
overabundant with respect to what observed in real GCs, guarantees
enough statistics to limit stochastic noise in the results.  We assume
the particles follow a King \citeyearpar{king66} model distribution
with no primordial mass segregation. In order to explore the effect of
different concentrations, we have chosen two different values of the
King central dimensionless potential: $W_0 = 5$ and $W_0 = 8$.

With the aim of exploring how cluster dynamics depends on the
  presence and content of heavy objects, and to isolate the effect of
  BHs from that of NSs, for each of the adopted $W_0$ values we ran
  five simulations by varying, in the initial conditions, the
  percentage of DRs retained within the system: $(i)$ in simulations
  $S^{W0}_0$ (with $W_0=5, 8$) no DRs have been retained (hence, BSSs
  are the most massive objects in the cluster); $(ii)$ in runs
  $S^{W0}_{10}$ and $S^{W0}_{30}$ we assumed that $10\%$ and $30\%$,
  respectively, of NSs have been retained, while all BHs have been
  ejected; $(iii)$ in simulations $S^{W0}_{10\bullet}$ and
  $S^{W0}_{30\bullet}$ we assumed that $10\%$ and $30\%$,
  respectively, of NSs and BHs have been retained.  The number of NSs,
  BHs, and BSSs in the initial conditions of each run are summarized
  in Table \ref{run_table}.

\section{Results}
\label{results}

\subsection{Evolution of the Lagrangian Radii}
\label{lagr} 
\begin{figure}[h!]
\centering \includegraphics[height=18.0cm,
  width=18.0cm]{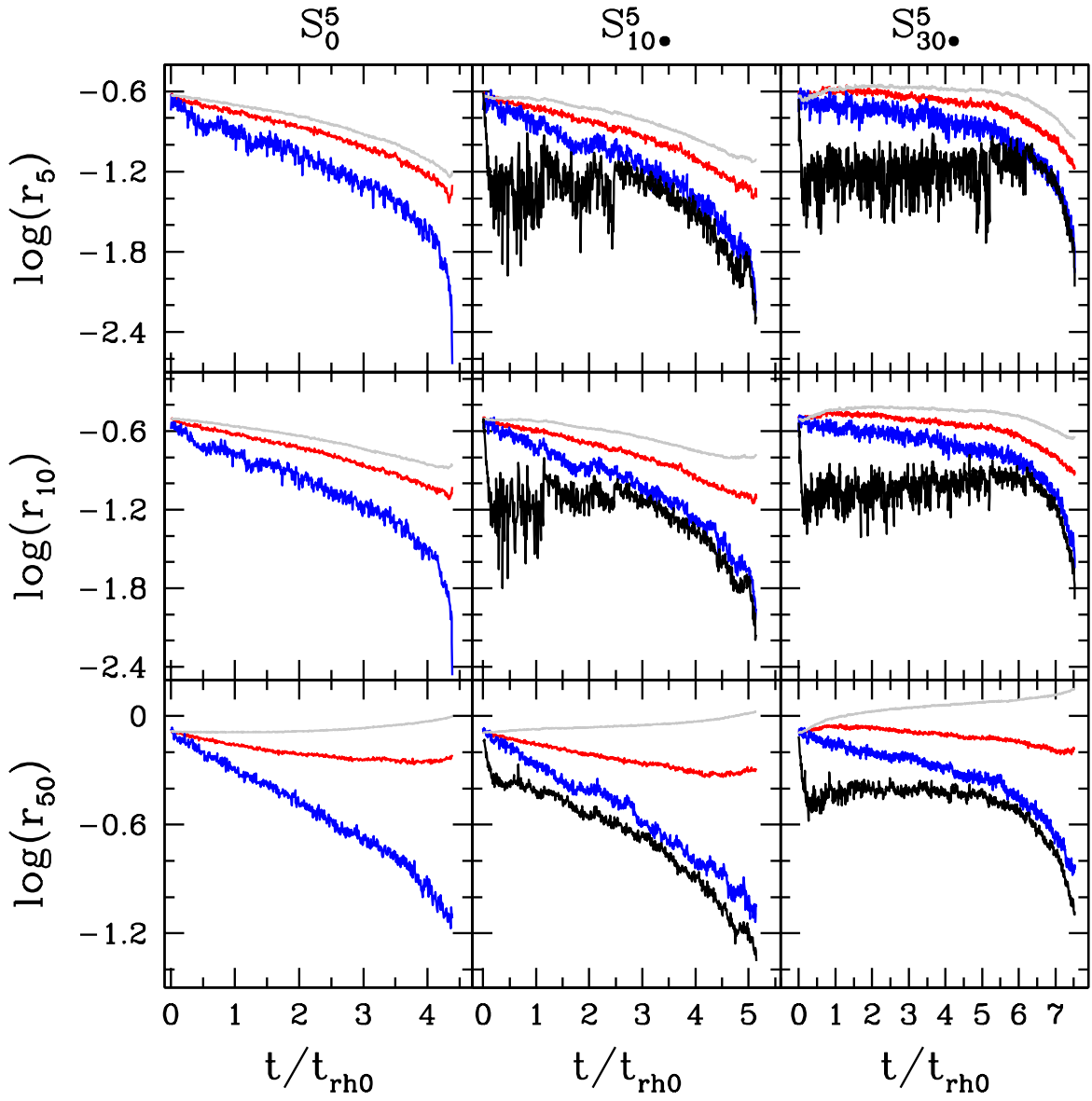}
\caption{Evolution of the Lagrangian radii containing $5\%, 10\%$ and
  $50\%$ (top, central and bottom panels, respectively) of the
  relative number of DRs (black), BSSs (blue), REF stars (red), and
  particles of any mass (grey), for the three runs with $W_0 = 5$ 
    that include BHs. Time is normalized to the initial half-mass
  relaxation time $t_{\rm rh0}$ of each run (see Table\ref{run_table}).}
\label{fig_lag5}
\end{figure}

\begin{figure}[h!]
\centering \includegraphics[height=18.0cm,  width=18.0cm]{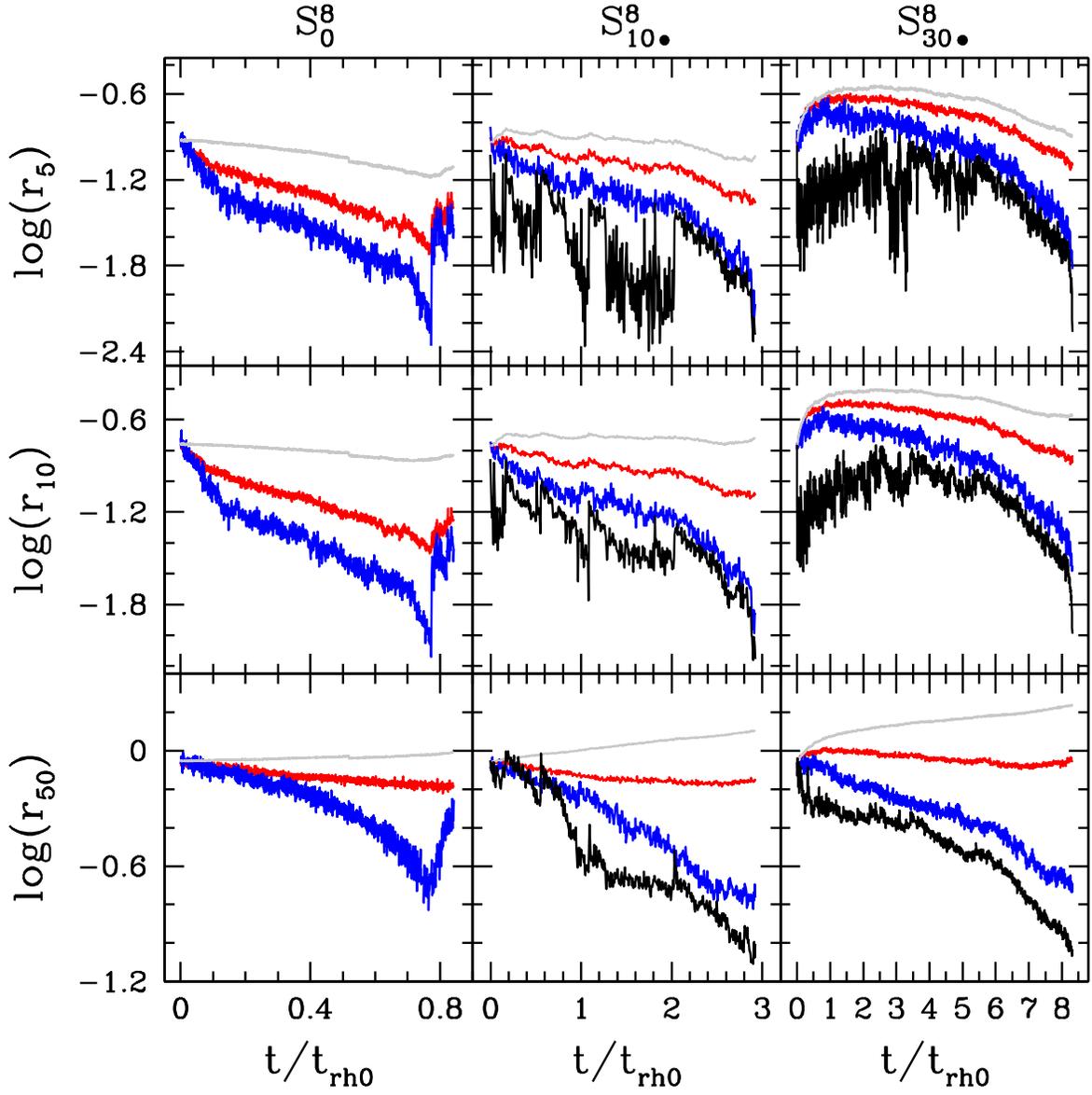}
\caption{Same as Figure \ref{fig_lag5} for the runs with $W_0 = 8$ that include BHs.}
\label{fig_lag8}
\end{figure}

We start the presentation of our results with a brief description of
the evolution of the structural properties of our simulated clusters,
and the effects of the presence of DRs. Figures \ref{fig_lag5} and
\ref{fig_lag8} show the evolution of the $5\%$, $10\%$ and $50\%$
number Lagrangian radii of particles belonging to different
populations in the simulations with no DRs and in those also
  including BHs, for $W_0 = 5$ and $W_0 = 8$, respectively.  In
particular, we compare the evolution of the Lagrangian radii of BSSs
(i.e., all the particles with mass $m = 1.2 \ M_\odot$; blue lines),
of what we call {\itshape{reference}} population (REF), corresponding
to all the particles with masses between 0.75 and $0.84 \ M_\odot$
(red lines), and of the overall system, including all particles
irrespective of their masses (grey lines). Moreover, for the runs with
BHs, we also show in black the evolution of the DR population
($m > 1.4 \ M_\odot$).  The figures clearly show that the cluster
dynamical evolution is highly affected by the presence and amount of
DRs, and also depends on the initial concentration of the
system. Quite interestingly, this is true not only in the innermost
cluster regions (as sampled by the $5\%$ and $10\%$ Lagrangian radii),
but the effects also extend much outward, with large differences even
at radial distances including $50\%$ of the populations.

The analysis of the results obtained for $W_0=5$ (Fig. \ref{fig_lag5})
shows that if no DRs are present (left-hand column) BSSs drive the
cluster toward CC. This is indeed expected (see also
\citealp{miocchi+15}), since in that case, BSSs are the most massive
objects within the system.  The time evolution of the Lagrangian
  radii in the cases where only NSs are retained (namely, simulations
  $S^5_{10}$ and $S^5_{30}$) is very similar to that with no DRs
  (thus, we provide no explicit figures for these runs), since the
  large majority of NSs has masses comparable to that of BSSs. Hence,
  the main effect of including a DR population made of NSs only is
  that the collapse is driven by these objects and BSSs segregate just
  at a slightly slower rate than in the case with no DRs at all.  In
the $S^5_{10\bullet}$ and even more in the $S^5_{30\bullet}$
simulations, instead, DRs undergo a rapid decoupling from the other
populations, forming a subsystem that quickly sinks toward the centre
(see the black curves in the central and the right columns of
Fig. \ref{fig_lag5}). Clearly this behaviour is due to the
  subsystem of BHs, which have masses significantly larger than BSSs
  and NSs. Analogous results have been found and discussed also by
\citet{sigurdsson+hernquist93} and \citet{kulkarni+93}, and again very
recently by \citet{banerjee+10} and \citet{breen+heggie13b}.
Interestingly, the effect of BHs on the time evolution of the
BSS and REF Lagrangian radii is negligible if a retention fraction of
only 10\% is assumed. In fact, their value in the $S^5_{10\bullet}$
run is almost the same, at any fixed value of time $t/t_{\rm rh0}$, as
in the $S^5_0$ (and the $S^5_{10}$) case.  In the $S_{10\bullet}^5$
simulation, after the initial phase of rapid DR decoupling, the inner
Lagrangian radii of these heavy objects stay approximately constant up
to $\sim 2.5 \ t_{\rm rh0}$, then decrease steadily in time, closely
followed by the Lagrangian radii of BSSs, which therefore start to
participate in driving the overall cluster evolution, similarly to
what happens in the $S_0^5$ run.  Instead, for $f_{\rm DR}= 30\%$ the
effect played by BHs is much stronger. After the initial
decoupling from the rest of the system, DRs evolve at almost constant
Lagrangian radii for approximately 6 initial relaxation times, while
the other stellar components migrate much more gently inward: the
Lagrangian radii of all populations are systematically larger, at the
same evolutionary time, with respect to what observed in the
$S_{10\bullet}^5$ case, and the difference between the Lagrangian
radii of the BSS and the REF populations is smaller.  The Lagrangian
radii of BSSs start to approach those of DRs at $\sim 6 \ t_{\rm
  rh0}$, while the same happens much earlier in the $S_{10\bullet}^5$
run. The overall system reacts with a continuos expansion of $r_{50}$
(grey line in the bottom right panel). The presence and the amount of
BHs in the system has a strong impact on the rate at which the
dynamical evolution of BSSs and REFs proceeds: in particular, the
evolution of the level of mass segregation of the BSS population is
increasingly inhibited and delayed as the adopted DR retention
fraction increases. In fact, while the time of CC of the visible component is $\sim
  4.4 \ t_{\rm rh0}$ if no DRs or only NSs are present, in the simulations including
also BHs, it increases to $\sim 5.2$ for a 10\% DR retention fraction, and
further to $\sim 7.5$ for $f_{\rm
  DR}=30\%$. These behaviours can be explained as an effect of
dynamical heating due to the population of BHs, which inhibits
mass segregation.  It is interesting to note that the inibhition of
the mass segregation, due to the heating from a BH subsystem,
resembles what \citet{baumgardt+04}, \citet{trenti+07} and
\citet{gill+08} have found in the presence of an intermediate-mass
black hole (IMBH). A recent study by \citet{peuten+16} has also
  shown that the lack of segregation observed in NGC 6101 might be due
  to a population of BHs, a result consistent with our findings.

Similar general comments as above also apply to the case of a much
more concentrated system ($W_0=8$, in place of $W_0=5$), where the
impact of DRs including a population of BHs is even more
apparent (see Fig. \ref{fig_lag8}). Overall the time evolution is much
faster than in the $W_0=5$ case, indicating that increasing the
cluster concentration accelerates the dynamical evolution of the
system. If DRs are absent, the inner Lagrangian radii ($r_5$ and
$r_{10}$) of both BSSs and REF stars rapidly decrease in time until the
CC (the rapid fluctuation in $r_5$ and $r_{10}$ are simply a
consequence of the ejection of BSS binaries), while $r_{50}$ of the
REF population remains almost constant during the entire evolution.
If NSs and BHs are included (central and right-hand columns) all
populations segregate much more slowly and the difference between the
BSS and the REF Lagrangian radii is smaller at any fixed evolutionary
times. In the $S_{30\bullet}^8$ run, DRs rapidly decouple from the
rest of the system, then their inner Lagrangian radii expand
significantly and start to re-contract only after $\sim 4 \ t_{\rm
  rh0}$, closely followed by BSSs.

Hence, according to what expected, massive objects (especially
  stellar-mass BHs) are found to play a fundamental role in the
cluster dynamics, driving the CC of visible stars and determining its
timescale.

\subsection{Cumulative radial distributions}
\label{cumu}
The time evolution of the Lagrangian radii discussed in the previous
section provides precious information about the processes of
segregation and expansion of various stellar populations. Although our
simulations are still idealized and not meant to be directly compared
to observational data, it is important to identify and follow the
evolution of mass segregation indicators that can be more easily
adopted in observational studies of the segregation of BSS and REF
populations. Many previous studies have shown
that different GCs are characterized by different cumulative radial
distributions, corresponding to different levels of BSS segregation in
the central regions (e.g., \citealp{ferraro+04}). Here we therefore
study how the cumulative radial distributions of BSSs and REFs depend
on the simulated cluster properties. In order to highlight the inner
distance scale, where the effect of DF is the most evident, we choose
to express the radial distance from the cluster centre in logarithmic units.
Since in all simulations we assume no initial mass segregation (see
Sect. \ref{Nbody}), at $t=0$ all populations are perfectly mixed and
the BSS and the REF distributions are superimposed.
However, as time increases, BSSs migrate toward the cluster centre
more rapidly than the REF population (see Figs. \ref{fig_lag5} and
\ref{fig_lag8}) and the two corresponding cumulative radial
distributions start to separate from each other.

\begin{figure}[h!]
\centering \includegraphics[height=18.0cm, width=18.0cm]{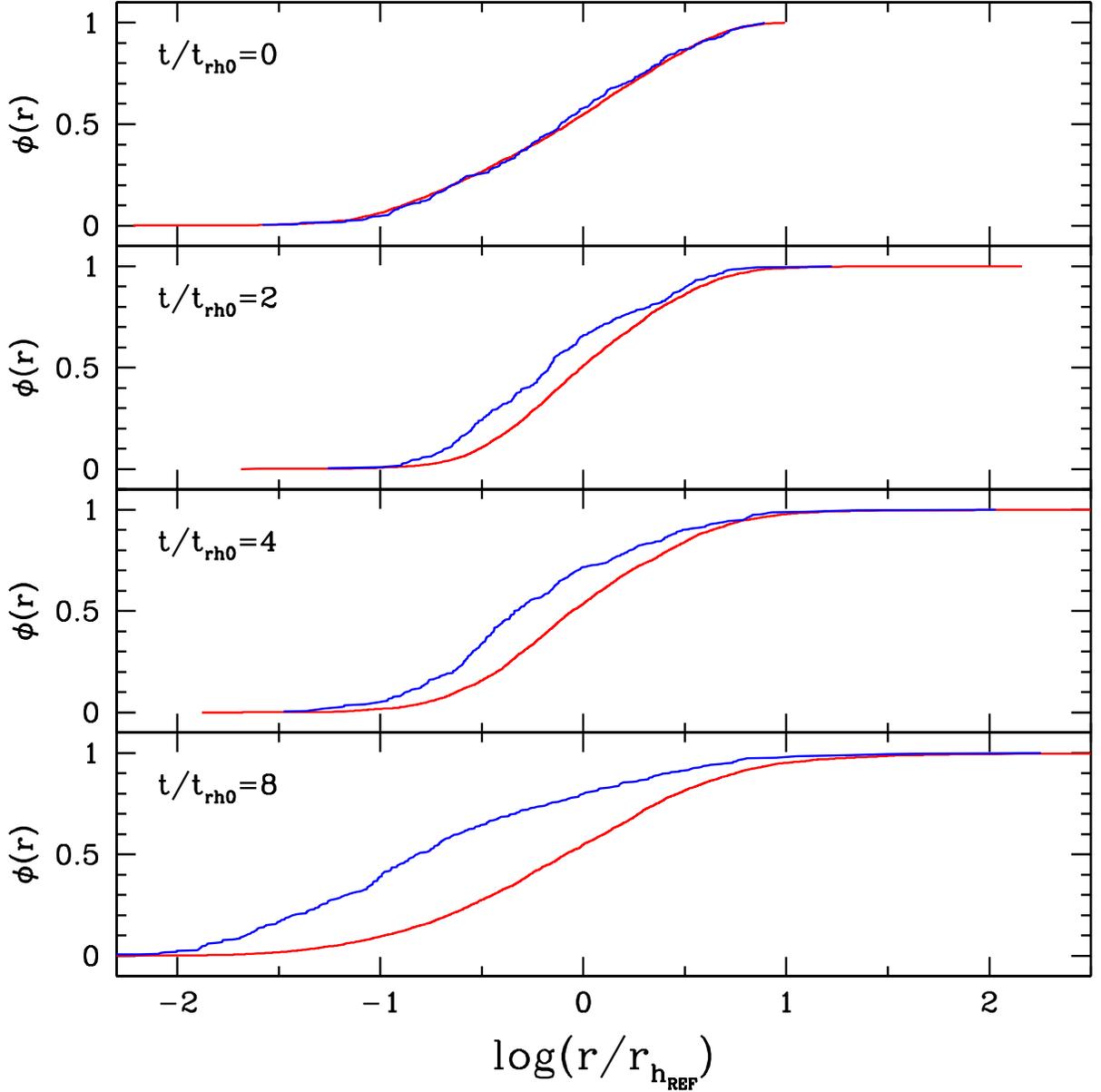}
\caption{From top to bottom, time evolution of the cumulative radial
  distributions of BSSs (blue lines) and REF stars (red lines), for
  the $S^8_{30\bullet}$ simulation. The radial scale is logarithmic,
  with the radius normalized to the half-mass radius of the REF
  population measured at any considered evolutionary time (see
  labels). At $t=0$ the two populations are perfectly mixed and their
  cumulative radial distributions superimposed. For increasing time,
  the two distributions become more and more separated due to the
  effect of mass segregation that preferentially segregates the
  heavier objects (BSSs) toward the clustre centre. The same
  qualitative trend is observed in all simulations.}
\label{cumu_w8}
\end{figure}

For illustrative purposes, Figure \ref{cumu_w8} shows the cumulative
radial distributions of BSSs and REFs (blue and red lines,
respectively) for the $S^8_{30\bullet}$ run, at four different evolutionary
times (normalized to the initial half-mass relaxation time of the run;
see labels in the figure). For the sake of comparison, the radial distance
is expressed in units of the half-mass radius of the REF population at
the considered time. A notable feature in the figure is that the
separation between the blue and the
red lines increases with time, with the BSS population always being
more centrally segregated (i.e., with a steeper cumulative
distribution) than the REF stars. This is due to DF that
preferentially affects the heavier component (BSSs, with respect to
REFs), making these objects more rapidly migrate toward the cluster
centre.

\begin{figure}[h!]
\centering \includegraphics[height=18.0cm, width=18.0cm]{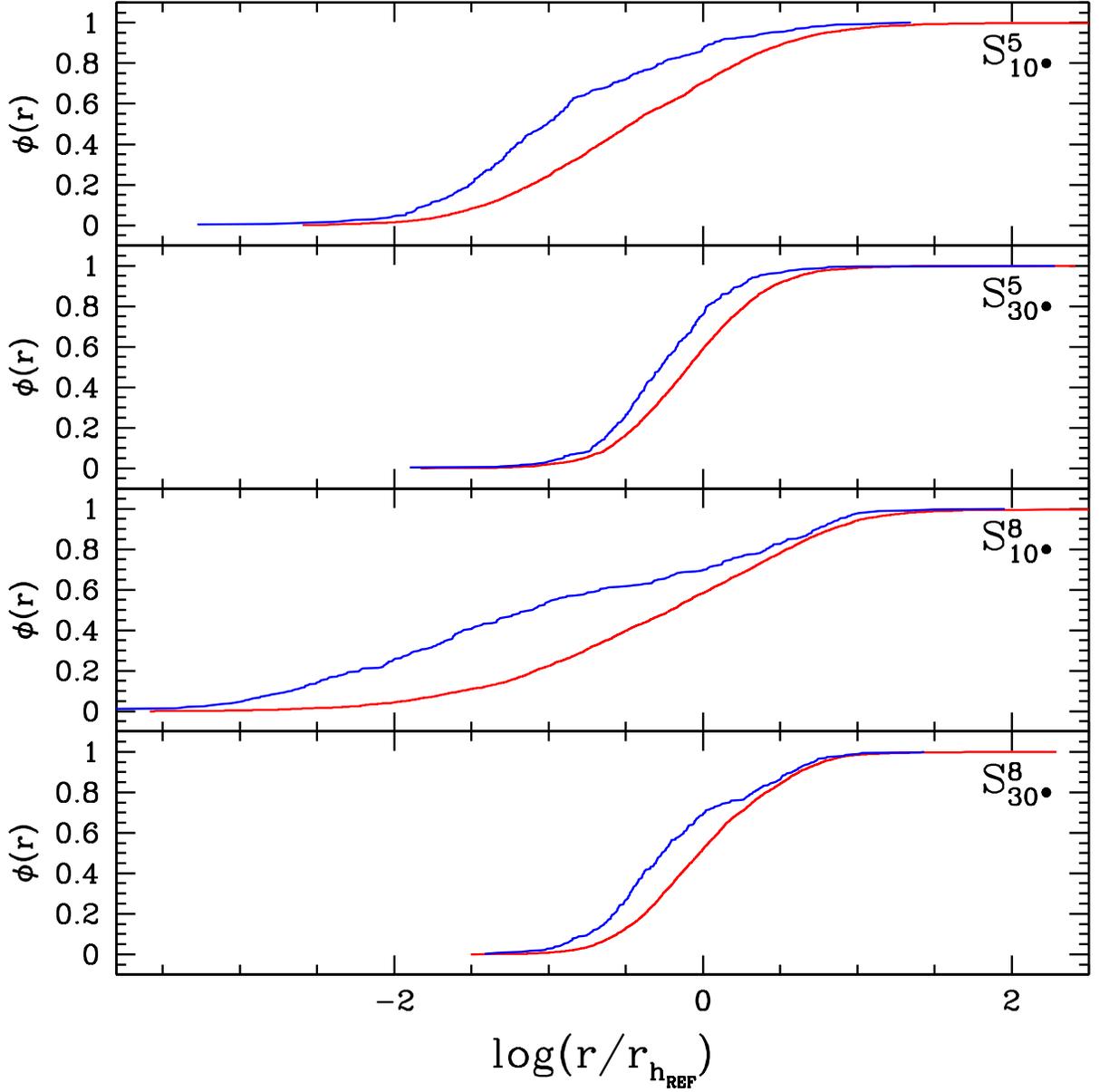}
\caption{Comparison among the cumulative radial distribution of BSSs
  (blue lines) and REF stars (red lines) at the same evolutionary time
  ($t=2.9 \ t_{\rm rh0}$) in the four runs including BHs (see
  labels).  The highest central concentration of BSSs (testified by
  both the lowest inner radius and the largest separation between the
  two distributions) is found for the $S^8_{10\bullet}$ run (which
  shows the fastest evolution: cfr with Figs. \ref{fig_lag5}). The
  smallest central segregation of BSSs is observed in the
  $S^5_{30\bullet}$ cluster (which, in fact, shows the slowest
  dynamical evolution).}
\label{cfr_cumu}
\end{figure}

Figure \ref{cfr_cumu} compares the cumulative radial distributions
obtained for the four models including BHs ($S^5_{10\bullet}$,
  $S^5_{30\bullet}$, $S^8_{10\bullet}$ and $S^8_{30\bullet}$) at a
fixed evolutionary time, $t=2.9 \ t_{\rm rh0}$ (when the
$S_{10\bullet}^8$ simulation stops). In line with what discussed in
the previous sections, we find that BSSs are more centrally segregated
for the lowest DR retention fraction (at fixed value of $W_0$) and for
the largest cluster concentration (at fixed DR retention fraction). In
particular, the accelerating effect on mass segregation due to a
larger cluster concentration is clearly dominant in the case of a
small population of DRs ($10\%$ retention fraction), while it is
almost cancelled or even overcome if DRs become sufficiently
numerous. In fact, for a fixed $30\%$ retention fraction the
cumulative radial distributions in the $W_0=5$ and $W_0=8$ cases are
quite alike, with the latter becoming different from zero at only
slightly lower radii than in the $W_0=5$ case (compare panels labeled
with $S^5_{30\bullet}$ and $S^8_{30\bullet}$ in
Fig. \ref{cfr_cumu}). Such an effect is even more apparent from the
comparison of the $S^5_{10\bullet}$ and $S^8_{30\bullet}$ radial
distributions: in spite of a smaller concentration (and thanks to a
lower fraction of DRs) both stellar populations are more concentrated
and the two cumulative distributions are more separated in the former
case ($S^5_{10\bullet}$).

\subsection{A new indicator of BSS segregation: $A^+$}
\label{a+}
The results discussed above suggest that the separation between the
cumulative radial distributions can be used to measure the level of
BSS central segregation with respect to a lighter cluster population
taken as reference. We quantitatively define this new indicator as the
area between the BSS and the REF cumulative radial distributions in
the $\phi(r)-\log(r/rh_{\rm REF})$ plane, and we name it $A^+$.

For perfectly mixed populations (as it is the case at $t=0$ in our
models), such a parameter must be equal to zero.  Then, $A^+$ is
expected to become positive because the effect of DF is stronger on
BSSs (heavier) than on the (lighter) REF stars, and the cumulative
radial distributions of the two populations therefore start to
separate from each other. According to the scenario presented by
\citet{ferraro+12}, as time passes, the value of $A^+$ is expected to
increase progressively, since BSSs orbiting at increasingly larger
distances from the cluster centre sink to the bottom of the potential
well and the increase of A+ with time corresponds to the formation and
the growth of the central peak of the normalized BSS distribution.

\begin{figure}[h!]
\centering
\includegraphics[height=18.0cm, width=18.0cm]{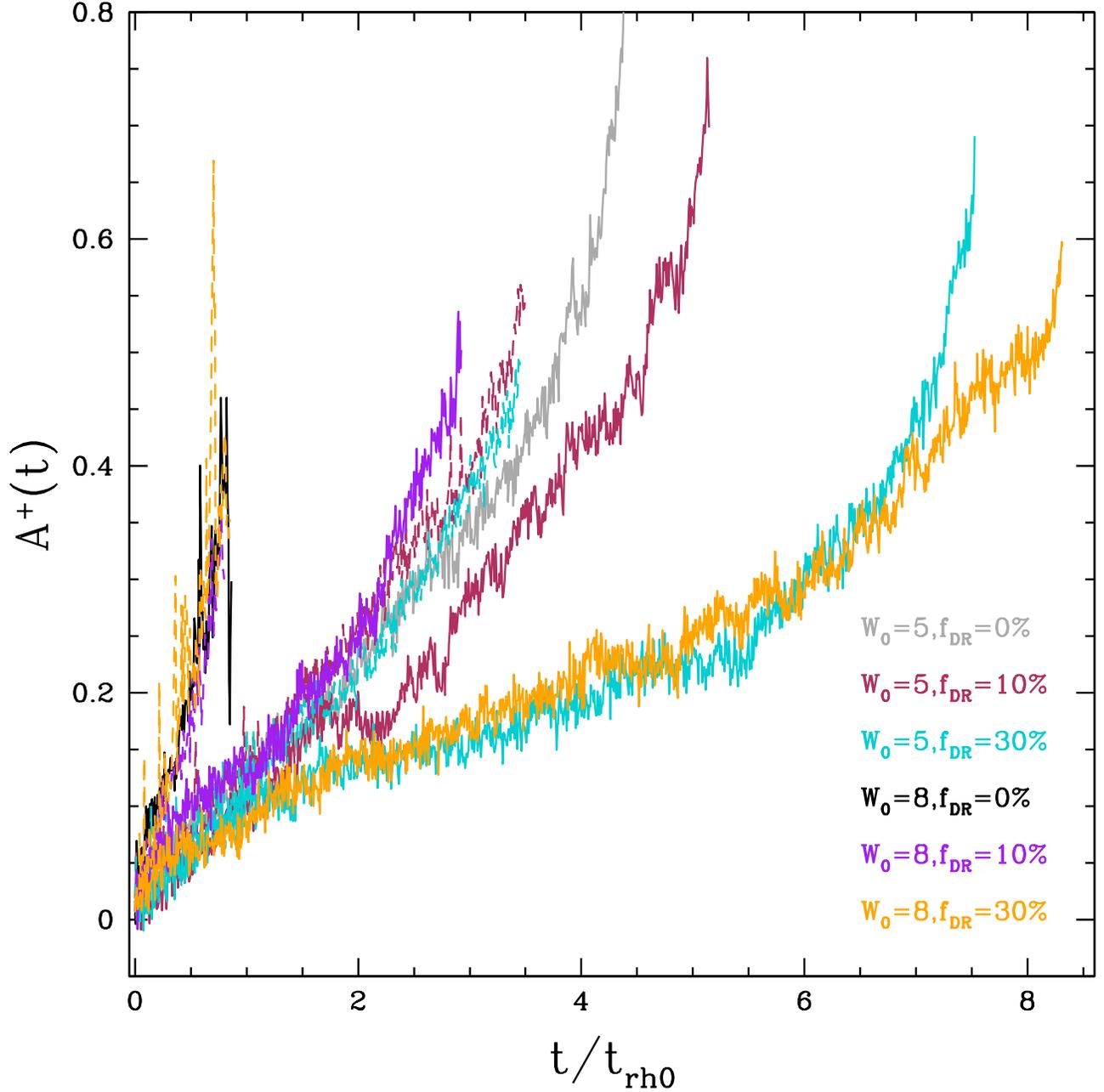}
\caption{Evolution of $A^+$ as a function of the time normalized
    to $t_{\rm rh0}$ in the all our models: simulations with no DRs
    are plotted in grey and black for the $W_0=5$ and $W_0=8$ cases,
    respectively; the dashed lines refer to models with only NSs,
    while the solid lines correspond to the simulations including both
    NSs and BHs (see the labels for the color code).  $A^+$ increases
  with time as expected for a reliable mass segregation and dynamical
  evolution indicator (see Sect. \ref{a+}).}
\label{fig_area_trh}
\end{figure}

Fig. \ref{fig_area_trh} shows that $A^+$ always increases with time
and confirms that this parameter is indeed a sensitive indicator of
the BSS sedimentation process. The figure confirms that NSs
  alone have a negligible impact on the rate of central BSS
  segregation: in fact, for a given value of $W_0$, there is no
  significant difference between the models with different percentages
  of such objects (see, for instance, the red and the cyan dashed
  lines: $S^5_{10}$ and $S^5_{30}$ runs, respectively), nor between
  these models and those with the same $W_0$ but no DRs (compare the
  lines above with the grey one, $S^5_0$, in the figure). Instead, BHs
  clearly appear to play a crucial role, significanlty slowing down
  the evolution of the system (see colored solid lines in the
  figure). In particular, we note that for a given concentration
($W_0$) and time ($t/t_{\rm rh0}$), the values of $A^+$ are larger for
models with a smaller BH retention. On the other hand, for a given
time and a given retention fraction, $A^+$ is larger for models with a
larger initial concentration. At any given value of $t/t_{\rm rh0}$,
the largest values of $A^+$ are found for models with the largest
concentration and the smallest BH retention fraction
($S^8_{10\bullet}$), while the smaller values of $A^+$ are found with
lower concentration models with larger BH retention fraction
($S^5_{30\bullet}$). In addition, the parameter is larger in the
$S^5_{10\bullet}$ case than in the $S^8_{30\bullet}$ run (at fixed
time), demonstrating that the (slowing) effect of a larger percentage
of BHs is stronger than the (accelerating) effect of a larger
cluster concentration. In our $W_0=8$ simulations all
  BHs are ejected from the system at the end of the run, while three
  and one BHs are still present at the end of the $S^5_{10\bullet}$
  and $S^5_{30\bullet}$ simulations, respectively (see also
  \citealp{heggie14, hurley+16} for examples of models in
which instead all or a large fraction of BHs are lost during the
cluster evolution). The gradual loss of
  the initial population of BHs allows the BSS to segregate and $A^+$
  to grow and reach values above 0.4-0.5 with a rate that is slower
  for increasing fractions of BHs.
The simulations presented here serve the purpose of illustrating the
general effect of a population of black holes on the process of
segregation of BSS and the extent to which this effect depends on the
fraction of BHs retained. The  time evolution of $A^+$ might differ
for clusters with different fractions of BHs.
In particular, should the initial fraction of BHs be larger and/or the
cluster structural properties allow to retain a larger fraction BHs
than those adopted here (see e.g., \citealp{morscher+15, chatterjee+16,
peuten+16}), the rate of BSS segregation (and
therefore the growth of $A^+$) could be slowed down and delayed more
than what shown in the few illustrative cases we have explored.

All these properties are consistent with the conclusions of our
discussion about the segregation process in terms of evolution of the
Lagrangian radii (see Sect. \ref{lagr}) and show that the new
parameter $A^+$ is a reliable indicator of the cluster dynamical
evolution as traced by the BSS population.

Fig. \ref{fig_area_trh} also shows that the time dependence of $A^+$
is characterized by two main regimes: an initial, slower phase,
followed by a steeper trend toward the end of the simulations, at
times approaching the CC time. This effect is more evident if BHs
  are included: the evolutionary times when the curves change regime
are $t\approx 4.2$, 5.5, 2 and $6.2 \ t_{\rm rh0}$ in the
$S_{10\bullet}^5$, $S^5_{30\bullet}$, $S^8_{10\bullet}$ and
$S^8_{30\bullet}$ runs, respectively. These epochs also correspond to
the moment when the time dependence of the inner BSS Lagrangian radii
also change slope, while those of the REF population evolve in a
slower, smoother way (see Figs. \ref{fig_lag5} and
\ref{fig_lag8}). The fact that $A^+$ shows a regime of faster time
dependence when the cluster is approaching the CC phase further
demonstrates that it is an indicator of the level of BSS segregation
and dynamical evolution of the system.

\begin{figure}[h!]
\centering
\includegraphics[height=18.0cm, width=18.0cm]{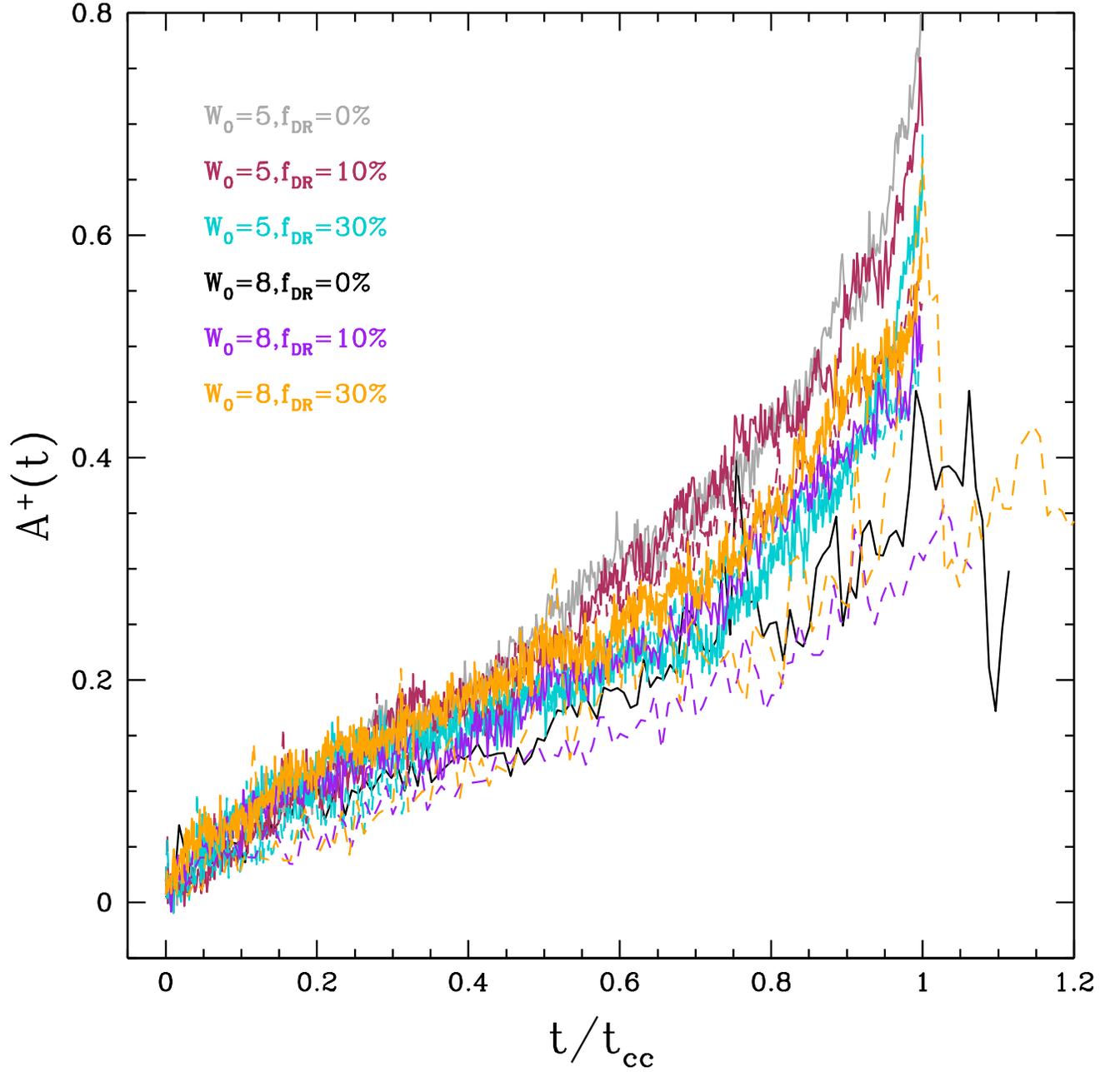}
\caption{Same as Fig. \ref{fig_area_trh}, but with the time
    normalized to the CC time of BSSs and REF stars in every model
    ($t_{\rm CC}$).}
\label{fig_area_tcc}
\end{figure}

Figure \ref{fig_area_tcc} shows the evolution of $A^+$ as a
  function of time normalized to the CC time of the BSS and REF
  populations ($t_{\rm CC}$). Interestingly, all the models fall on a
  relatively narrow band in this plane, irrespective of the amount and
  the mass of DRs they contain.  This suggests that $A^+$ might be
  used as an approximate indicator of how far a cluster is from CC of
  the visible component.

\section{Summary and conclusions}
\label{conclusions}
We performed a set of direct N-body simulations of GCs with $N \sim
10^5$ particles, with different initial concentrations ($W_0=5$ and
$W_0=8$), and admitting a population of BSSs and different
  fractions of NSs and BHs (including the case of systems with no
heavy DRs at all). BSSs are modelled as single particles of $1.2
M_\odot$.  The simulations have been used to investigate how a
population of heavy objects affects mass segregation and the dynamical
evolution of a cluster, and what can be learned from BSSs about these
processes.

We have shown that the segregation of BSSs and of the most massive
main sequence stars is significantly slowed down by the presence of
BHs. This is due to the dynamical heating effect of these heavy
objects that rapidly decouple from the other components, form a
centrally concentrated sub-system and inhibit mass segregation of the
less massive components. The effect of BHs stronger than that
of concentration: in fact, in spite of a larger value of $W_0$, the
mass segregation process of currently visible stars (with masses
  $\le 0.8 M_\odot$) in the $S^8_{30\bullet}$ run is much slower than
that of the $S^5_{10\bullet}$ simulation.

Because of their larger masses, BSSs are expected to be more affected
by the mass segregation process and to be more centrally segregated
than the REF population.  This has been indeed observed in several
previous studies, also showing that the level of BSS segregation (and
thus the relative separation between the BSS and the REF cumulative
radial distributions) varies from cluster to cluster (see, e.g.,
Figure 2 in \citealp{ferraro+03}). In the framework of the dynamical
clock \citep{ferraro+12}, this is the result of DF, which favors the
sedimentation of BSSs toward the cluster center, causing the formation
of a central peak in the normalized BSS radial distribution. Here we
have proposed the parameter $A^+$ as a new tool to measure this
sedimentation level.  We found that $A^+$ shows a clear increasing
trend with time that well reflects the evolution of the Lagrangian
radii observed in the various runs, with analogous dependences on the
cluster concentration and the DR retention fraction. As the system
evolves and loses part of the initial population of BHs, the
level of BSS segregation grows at an increasing rate and diverge from
that of the REF stars. As a consequence, the time dependence of $A^+$
shows a change in the slope.  This demonstrates that $A^+$ is indeed a
very good tracer of BSS segregation, and can reveal both how advanced
BSS segregation is and the extent to which the presence of a DR
population has inhibited it. In all the considered simulations, the
parameter assumes comparable values at any fixed fraction of the
  CC time (see Fig. \ref{fig_area_tcc}). Hence, it seems to be a good
  indicator of the time remaining to the CC of the visible component,
  reasonably irrespective of the initial concentration and DR content
  of the system.

\begin{figure}[h!]
\centering
\includegraphics[height=18.0cm, width=18.0cm]{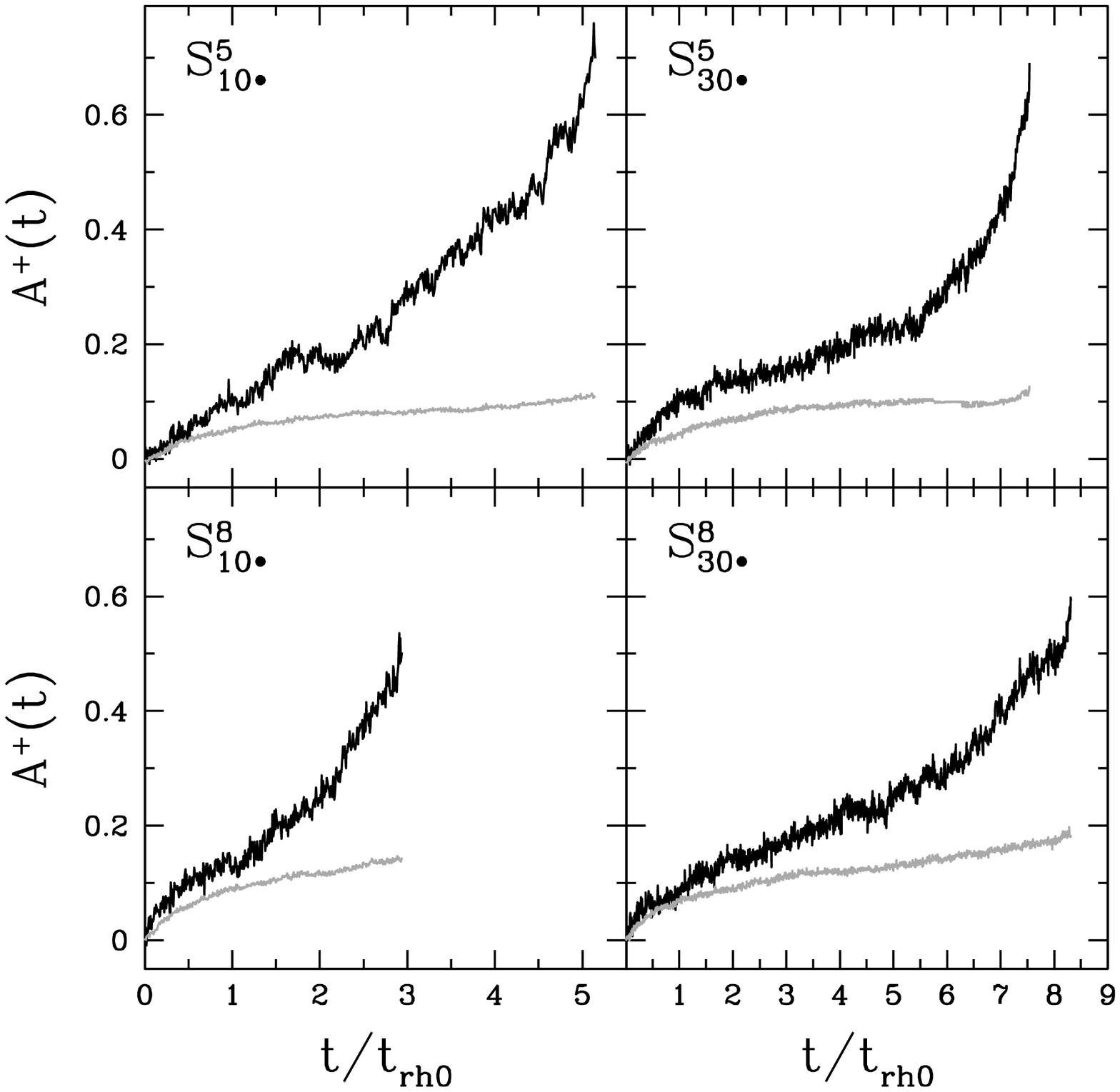}
\caption{Time evolution of the parameter $A^+$ computed from two
  different pairs of populations for our set of simulations (see
  labels). The black curves are the same as the solid colored lines in
  Fig. \ref{fig_area_trh} (obtained from the BSS and the REF
  populations in the models with BHs)
  while the grey lines mark the values of the area between the
  cumulative radial distributions of REF stars and $m = 0.4 M_{\odot}$
  particles.}
\label{area_low}
\end{figure}

In order to illustrate how the strength of the mass segregation
depends on the populations chosen, in Fig. \ref{area_low} we compare
the time evolution of the $A^+$ parameter calculated for BSSs and REF
stars (black curves, the same as those plotted in colors in
Fig. \ref{fig_area_trh}) and the time evolution of the $A^+$ parameter
calculated using REF and $m=0.4 M_{\odot}$ particles (grey curves). In
the former case the mass ratio between the two populations is $1.5$,
while it is larger (equal to 2) in the latter. While in both cases
$A^+$ increases with time, the effect is much stronger in the
former. This is because, in spite of a smaller mass ratio between the
former populations (BSSs and REFs), the relative effect of mass
segregation on these components is stronger than that on REF and $0.4
M_{\odot}$ stars.  This behaviour suggests that BSSs are more powerful
observational tracers of dynamical evolution than, for instance, main
sequence stars.

Although our simulations are still idealized, they illustrate the
general expected evolution of the $A^+$ parameter, and its dependence
on the evolutionary stage of a cluster and its DR content. A larger
exploration of the parameter space, including a wider grid of values
for $W_0$ and $f_{\rm DR}$, a population of primordial binaries and
the effect of a host galaxy tidal field, is needed before drawing more
quantitative conclusions. These aspects will be studied in future
investigations. On the observational side, we just provided the
  first empirical determination of $A^+$ in a sample 25 of Galactic
  GCs, demonstrating that it shows tight correlations both with the
  radial position of the minimum of their BSS distribution and with
  the cluster central relaxation time (Lanzoni et al. 2016, ApJ
  submitted), further confirming that it is a powerful indicator of GC
  dynamical evolution.

\section*{Acknowledgements}
This research is part of the project Cosmic-Lab (web site:
http://www.cosmic-lab.eu) funded by the European Research Council
(under contract ERC-2010-AdG-267675).  EA thanks the {\it Marco Polo
  Project} of the Bologna University and the Indiana University for
the hospitality during his stay in Bloomington, where part of this
work was carried out.  This research was supported in part by Lilly
Endowment, Inc., through its support for the Indiana University
Pervasive Technology Institute, and in part by the Indiana METACyt
Initiative. The Indiana METACyt Initiative at IU is also supported in
part by Lilly Endowment, Inc.  We warmly thank Dr. Jongsuk Hong for
stimulating discussions and his support in running the N-body
simulations.  We also thank the anonymous Referee for useful comments
that improved the presentation of the paper.

\newpage
\begin{table}
\begin{center}
\begin{tabular}{lcrrrrrr}
\hline
Run & $W_0$ & $f_{\rm DR}$ & $N_{\rm NS0}$ & $N_{\rm BH0}$ & $N_{\rm BSS0}$ & $N_{\rm TOT}$ & $t_{\rm rh0}$ \\
  & & \% & & & & \\
\hline
\hline
$S^5_0$   & $5$ &   $0$ & $0$ &  $0$ & $300$ & $10^5$ & $1330.5$ \\
$S^5_{10}$ & $5$ & $10$ & $72$ & $0$ & $300$ & $99161$ & $1325.4$ \\
$S^5_{30}$ & $5$ & $30$ & $206$ & $0$ & $300$ & $99295$ & $1326.9$ \\
$S^5_{10\bullet}$ & $5$ & $10$ & $72$ & $19$ & $300$ & $99180$ & $1325.6$ \\
$S^5_{30\bullet}$ & $5$ & $30$ & $206$ & $67$ & $300$ & $99362$ & $1328.8$ \\
$S^8_0$   & $8$ &   $0$ & $0$ & $0$ & $300$ & $10^5$ & $1471.7$ \\
$S^8_{10}$ & $8$ & $10$ & $71$ & $0$ & $300$ & $99183$ & $1465.3$ \\
$S^8_{30}$ & $8$ & $30$ & $200$ &  $0$ & $300$ & $99313$ & $1466.9$ \\
$S^8_{10\bullet}$ & $8$ & $10$ & $71$ & $17$ & $300$ & $99202$ & $1470.6$ \\
$S^8_{30\bullet}$ & $8$ & $30$ & $200$ &  $65$ & $300$ & $99379$ & $1467.8$ \\
\hline
\end{tabular} 
\end{center}
\caption{Initial conditions of the N-body simulations. For each run,
  the table lists the adopted name (column 1), the intial value of the
  King dimensionless potential $W_0$ (column 2), the initial retention
  fraction of DRs (column 3), the total number of DRs, BHs, BSSs and
  particles of any mass at $t=0$ (columns 4--7), the initial half-mass
  relaxation time expressed in N-body units (column 8).}
\label{run_table}
\end{table}

\newpage

\newpage

\newpage

\newpage

\newpage

\newpage

\newpage


\newpage
\begin{thebibliography}{99}

\bibitem[\protect\citeauthoryear{Alessandrini et al.}{2014}]{alessandrini+14} 
Alessandrini, E., Lanzoni, B., Miocchi, P., Ciotti, L., Ferraro, F. R., 
2014, \apj, 795, 169.

\bibitem[\protect\citeauthoryear{Banerjee et al.}{2010}]{banerjee+10} 
Banerjee, S., Baumgardt, H., Kroupa, P., 
2010, \mnras, 402, 371.

\bibitem[\protect\citeauthoryear{Baumgardt et al.}{2004}]{baumgardt+04} 
Baumgardt, H., Makino, J., Ebisuzaki, T., 
2004, \apj, 613, 1143.

\bibitem[\protect\citeauthoryear{Beccari et al.}{2011}]{beccari+11} 
Beccari, G., Sollima, A., Ferraro, F. R., Lanzoni, B., Bellazzini, M., De
Marchi, G., Valls-Gabaud, D., Rood, R. T., 
2011, \apj, 737, 3.

\bibitem[\protect\citeauthoryear{Breen \& Heggie}{2013a}]{breen+heggie13a} 
Breen, P. G., Heggie, D.,
2013a, \mnras, 432, 2779.

\bibitem[\protect\citeauthoryear{Breen \& Heggie}{2013b}]{breen+heggie13b} 
Breen, P. G., Heggie, D.,
2013b, \mnras, 436, 584.

\bibitem[\protect\citeauthoryear{Binney \& Tremaine}{2008}]{binney+tremaine} Binney,
  J., Tremaine, S., 2008, Galactic Dynamics. Princeton University Press,
  Princeton, USA.

\bibitem[\protect\citeauthoryear{Chandrasekhar}{1943}]{chandrasekhar43} 
Chandrasekhar, S.,
1943, \apj, 97, 255.

\bibitem[Chatterjee et al.(2016)]{chatterjee+16} Chatterjee, S.,
  Rodriguez, C., L., \& Rasio, F., A.\ 2016, preprint (arXiv:1603.00884) 

\bibitem[\protect\citeauthoryear{Dalessandro et al.}{2008}]{dalessandro+08} 
Dalessandro, E., Lanzoni, B., Ferraro, F. R., Vespe, F., Bellazzini, M., Rood, R. T., 
2008, \apj, 681, 311.

\bibitem[\protect\citeauthoryear{Dalessandro et al.}{2012}]{dalessandro+12}
Dalessandro, E., Schiavon, R. P., Rood, R. T., Ferraro, F. R., Sohn, S. T., Lanzoni, B., O'Connell, R. W.,
2012, \aj, 144, 126.

\bibitem[\protect\citeauthoryear{Davies \& Hansen}{1998}]{davies+hansen98} 
Davies, M. B., Hansen, B. M. S.,
1998, \mnras, 301, 15.

\bibitem[\protect\citeauthoryear{Drukier}{1996}]{drukier96} 
Drukier, G. A.,
1996, \mnras, 280, 498.

\bibitem[\protect\citeauthoryear{Ferraro et al.}{1993}]{ferraro+93}
Ferraro, F. R., Fusi Pecci, F., Cacciari, C.,
1993, \aj, 106, 2324.

\bibitem[\protect\citeauthoryear{Ferraro et al.}{1997}]{ferraro+97} 
Ferraro, F. R., Paltrinieri, B., Fusi Pecci, F., Cacciari, C., Dorman, B.,
Rood, R. T., Buonanno, R., Corsi, C. E., Burgarella, D., Laget, M.,
1997, A\&A, 324, 915.

\bibitem[\protect\citeauthoryear{Ferraro et al.}{1999}]{ferraro+99} 
Ferraro, F. R., Paltrinieri, B., Rood, R. T., Dorman, B.,
1999, \apj, 522, 983.

\bibitem[\protect\citeauthoryear{Ferraro et al.}{2003}]{ferraro+03} 
Ferraro, F. R., Sills, A., Rood, R. T., Paltrinieri, B., Buonanno, R.,
2003, \apj, 588, 464.

\bibitem[\protect\citeauthoryear{Ferraro et al.}{2004}]{ferraro+04} 
Ferraro F. R.,Beccari G., Rood, R. T., Bellazzini M., Sills A., Sabbi E.,
2004 \apj, 603, 127. 

\bibitem[\protect\citeauthoryear{Ferraro et al.}{2006a}]{ferraro+06a}
Ferraro, F. R., Sabbi, E., Gratton R., Piotto, G., Lanzoni, B., Carretta,
E., Rood, R. T., Sills, A., Fusi Pecci, F., Moehler, S., Beccari, G.,
Lucatello, S., Compagni, N.,
2006a, \apj, 647, L53.

\bibitem[\protect\citeauthoryear{Ferraro et al.}{2006b}]{ferraro+06b}
Ferraro, F. R., Sollima, A., Rood, R. T., Origlia, L., Pancino, E., Bellazzini, M., 
2006b, \apj, 638, 433.

\bibitem[\protect\citeauthoryear{Ferraro et al.}{2012}]{ferraro+12}
Ferraro, F. R., Lanzoni, B., Dalessandro, E., Beccari, G., Pasquato, M., Miocchi, P., 
Rood, R. T., Sigurdsson, S., Sills, A., Vesperini, E., Mapelli, M., Contreras, R., 
Sanna, N., Mucciarelli, A.,
2012, \nat, 492, 393.

\bibitem[\protect\citeauthoryear{Fiorentino et al.}{2014}]{fiorentino+14}
Fiorentino, G., Lanzoni, B., Dalessandro, E., Ferraro, F. R., Bono, G., Marconi, M.,
2014, \apj, 783, 29.

\bibitem[\protect\citeauthoryear{Gill et al.}{2008}]{gill+08} 
Gill, M., Trenti, M., Miller, M. C., van der Marel, R., Hamilton, D.,
Stiavelli, M., 
2008, \apj, 686, 303.

\bibitem[\protect\citeauthoryear{Heggie}{2014}]{heggie14} 
Heggie, D.,
2014, \mnras, 445, 3435.

\bibitem[\protect\citeauthoryear{Hurley et al.}{2000}]{hurley+00} 
Hurley,  J. R., Pols, O. R., Tout, C. A., 
2000, \mnras, 315, 543.

\bibitem[\protect\citeauthoryear{Hurley et al.}{2002}]{hurley+02}
Hurley,  J. R., Pols, O. R., Tout, C. A., 
2002, \mnras, 329, 897.

\bibitem[\protect\citeauthoryear{Hurley et al.}{2016}]{hurley+16}
Hurley,  J. R., Sippel, A. C., Aarseth, S., Tout, C. A., 
2016, pasa, 33, 36.

Hurley et al. 2016 PASA, 33,36 (http://arxiv.org/abs/1607.00641)

\bibitem[\protect\citeauthoryear{Hut et al.}{1991}]{hut+91}
Hut, P., Murphy, B. W., Verbunt, F., 
1991, A\&A, 241, 137.

\bibitem[\protect\citeauthoryear{King}{1966}]{king66} 
King, I. R., 
1966, \aj, 71, 64.

\bibitem[\protect\citeauthoryear{Kroupa}{2001}]{kroupa01} 
Kroupa, P., 
2001, \mnras, 322, 231.

\bibitem[\protect\citeauthoryear{K{\"u}pper et al.}{2011}]{kupper+11} 
K{\"u}pper, A. H. W., Maschberger, Th., Kroupa, P., Baumgardt, H., 
2011, \mnras, 417, 2300.

\bibitem[\protect\citeauthoryear{Kulkarni et al.}{1993}]{kulkarni+93} 
Kulkarni, S. R., Hut, P., McMillan, S.,
1993, \nat, 364, 421.

\bibitem[\protect\citeauthoryear{Lanzoni et al.}{2007a}]{lanzoni+07a} 
Lanzoni, B., Sanna, N., Ferraro, F. R., Valenti, E., Beccari, G., Schiavon,
R. P., Rood, R. T., Mapelli, M., Sigurdsson, S., 
2007a, \apj, 663, 1040.

\bibitem[\protect\citeauthoryear{Lanzoni et al.}{2007b}]{lanzoni+07b} 
Lanzoni, B., Dalessandro, E., Ferraro, F. R., Mancini, C., Beccari, G.,
Rood, R. T., Mapelli, M., Sigurdsson, S., 
2007b, \apj, 663, 267.

\bibitem[\protect\citeauthoryear{Mapelli et al.}{2004}]{mapelli+04} 
Mapelli, M., Sigurdsson, S., Colpi, M., Ferraro, F. R., Possenti, A., Rood, R. T.; Sills, A.; Beccari, G.,
2004, \apjl, 605, L29.

\bibitem[\protect\citeauthoryear{Mapelli et al.}{2006}]{mapelli+06} 
Mapelli, M., Sigurdsson, S., Ferraro, F. R., Colpi, M., Possenti, A., Lanzoni, B., 
2006, \mnras, 373, 361.

\bibitem[\protect\citeauthoryear{Miocchi et al.}{2015}]{miocchi+15} 
Miocchi, P., Pasquato, M., Lanzoni, B., Ferraro, F. R., Dalessandro, E., Vesperini, E., Alessandrini, E., Lee, Y.-W.,
2015, \apj, 799, 44.

\bibitem[\protect\citeauthoryear{Mackey et al.}{2007}]{mackey+07} 
Mackey, A. D., Wilkinson, M. I., Davies, M. B., Gilmore, G. F., 
2007, \mnras, 379, L40.

\bibitem[\protect\citeauthoryear{Mackey et al.}{2008}]{mackey+08} 
Mackey, A. D., Wilkinson, M. I., Davies, M. B., Gilmore, G. F., 
2008, \mnras, 386, 65.

\bibitem[\protect\citeauthoryear{Moody \& Sigurdsson}{2009}]{moody+sigurdsson09} 
Moody, K., \& Sigurdsson, S., 
2009, \apj, 690, 1370.

\bibitem[\protect\citeauthoryear{Morscher et al.}{2013}]{morscher+13} 
Morscher, M., Umbreit, S., Farr, W. M., Rasio, F. A., 
2013, \apj, 763, 15.

\bibitem[\protect\citeauthoryear{Morscher et al.}{2015}]{morscher+15} 
Morscher, M., Bharath, P., Rodriguez, C., Rasio, F. A., Umbreit, S., 
2015, \apj, 800, 9.

\bibitem[\protect\citeauthoryear{Nitadori \& Aarseth}{2012}]{nitadori+aarseth12} 
Nitadori, K., \& Aarseth, S. J., 
2012, \mnras, 424, 545.

\bibitem[\protect\citeauthoryear{O'Leary et al.}{2006}]{oleary+06} 
O’Leary, R. M., Rasio, F. A., Fregeau, J. M., Ivanova, N., O’Shaughnessy, R.,
2006, \apj, 637, 937.

\bibitem[\protect\citeauthoryear{Pfahl et al.}{2002}]{pfahl+02} 
Pfahl, E., Rappaport S., Podsiadlowski P., 
2002, \apj, 573, 283.

\bibitem[\protect\citeauthoryear{Peuten et al.}{2016}]{peuten+16} 
Peuten, M., Zocchi, A., Gieles, M., Gualandris, A., Henault-Brunet, V.,
2016, \mnras, 462, 2333.

\bibitem[\protect\citeauthoryear{Podsiadlowski et al.}{2005}]{podsiadlowski+05} 
Podsiadlowski P., Pfahl, E., Rappaport S., 
2005, {\itshape{ASP Conference Series}}, 328.

\bibitem[\protect\citeauthoryear{Repetto et al.}{2012}]{repetto+12} 
Repetto S., Davies, M. B., Sigurdsson S., 
2012, \mnras, 425, 2799.

\bibitem[\protect\citeauthoryear{Sandage}{1953}]{sandage53} Sandage
 A. R., 1953, \aj, 58, 61.

\bibitem[\protect\citeauthoryear{Shara et al.}{1997}]{shara97} Shara, M. M.,
  Saffer R. A., Livio M., 1997, \apj, 489, 59.

\bibitem[\protect\citeauthoryear{Sigurdsson \& Hernquist}{1993}]{sigurdsson+hernquist93} 
Sigurdsson, S., Hernquist, L.,
1993, \nat, 364, 423.

\bibitem[\protect\citeauthoryear{Sippel \& Hurley}{2013}]{sippel+hurley13} 
Sippel, A. C., Hurley, J. R.,
2013, \mnras, 430, L30.

\bibitem[\protect\citeauthoryear{Trenti et al.}{2007}]{trenti+07} 
Trenti, M., Ardi, E., Shin, M., Hut, P., 
2007, \mnras, 374, 857.

\end{thebibliography}
\end{document}